\documentstyle[aps,prl,epsf]{revtex}
\def\be{\begin{equation}}
\def\ee{\end{equation}}
\def\ba{\begin{eqnarray}}
\def\ea{\end{eqnarray}}

\title{Reconstruction of the primordial Universe by 
a Monge--Amp\`ere--Kantorovich optimisation scheme}
\author{Roya Mohayaee$^1$, Uriel Frisch$^1$, 
Sabino Matarrese$^2$ and Andrei Sobolevski{\u\i}$^{1,3}$\,\,\,\,}
\address{
$^1$D\'epartement Cassini, Observatoire de la C\^ote d'Azur, 
BP 4229, 06304 Nice, Cedex 4, France\\
$^2$Dipartimento di Fisica "Galileo Galilei'' and INFN, Sezione di
Padova,
via Marzolo 8, I-35131 Padova, Italy\\
$^3$ Department of Physics, M V Lomonossov 
University, 119899-Moscow, Russia
}
\begin{document}
%Date {1st/Feb./2003}
%submitted to: Astronomy \& Astrophysics
\maketitle
\begin{abstract}
A method for the reconstruction of the 
primordial density fluctuation field
is presented.
Various previous approaches to this problem rendered {\it non-unique}
solutions. Here,
it is demonstrated that the initial positions of dark matter 
fluid elements, under the hypothesis
that their displacement is the gradient of a convex potential, can 
be reconstructed uniquely.
In our approach, the cosmological reconstruction 
problem is reformulated as an assignment problem
in optimisation theory. When tested against numerical simulations, 
our scheme yields excellent reconstruction on scales larger than a few 
megaparsecs.
\end{abstract}

%\keywords{Reconstruction, Primordial universe}}

%\titlerunning{Reconstruction of the primordial ...}

%\authorrunnng{}

%\begin{document}

%\maketitle

\section{Introduction}

The present distribution of galaxies brought to us by redshift surveys
indicates that the Universe on large scales exhibits a high degree of
clumpiness with coherent 
structures such as voids, great walls, filaments and clusters. 
The cosmic microwave background (CMB) explorers, however,
indicate that the Universe was highly homogeneous billions of years ago. 
When studying these data, among the questions that are of concern in 
cosmology are the initial conditions of the Universe and the dynamics 
under which it grew into its present form. CMB
explorers provide us with valuable knowledge into the initial conditions 
of the Universe, but the present distribution of the galaxies opens a
second, complementary window into its early state. 

Unraveling the properties of the early Universe from the present data is an
instance of the general class of {\it inverse problems} in physics. 
In cosmology this problem is frequently tackled in an
empirical way by taking a {\it forward approach}. 
In this approach, a cosmological model is proposed for the initial 
power spectrum of dark matter. Next, a particle presentation of the 
initial density field is made which provides the initial data 
for an N-body simulation which 
is run using Newtonian dynamics and is stopped at the present
time. Subsequently, a {\it statistical} comparison between the outcome 
of the simulation 
and the observational data can be made, assuming that a suitable {\it bias} 
relation exists between the distribution of galaxies and that of dark 
matter. If the statistical test is satisfactory then the
implication is that the initial condition was viable; otherwise one changes 
the cosmological parameters and goes through the whole
process again. This is repeated until one obtains a satisfactory statistical
test, affirming a good choice for the initial condition.

However, this inverse problem does not have to be necessarily dealt 
with in a forward manner. 
Can one fit the present distribution of the galaxies {\it exactly} rather 
than statistically and run it back in time to make the {\it reconstruction} 
of the primordial density fluctuation field?
Since Newtonian gravity is time-reversible, one would have 
been able to integrate the equations of motions
back in time and solve the reconstruction problem trivially, if in 
addition to their positions, the present velocities of the 
galaxies were also known.
As a matter of fact, however, the peculiar velocities of only a 
few thousands of galaxies are known out of the hundreds of thousands whose 
redshifts have been measured. Indeed, one goal of reconstruction is to 
evaluate the peculiar velocities of the galaxies and in this manner put 
direct constraints on cosmological parameters.

Without a second boundary condition, reconstruction
would thus be an infeasible task. Newton's equation 
of motion requires two boundary conditions, whereas, for
reconstruction so far we only have mentioned the present positions of
the galaxies. The second condition is the
homogeneity of the initial density field: as we go back in time 
the peculiar velocities of
the galaxies vanish.
Thus, contrary to the forward approach where one solves an {\it initial-value
problem}, in the reconstruction approach one is dealing 
with a {\it two-point boundary value problem}. In the former, one starts with the 
initial positions and velocities of the
particles and solves Newton's equations 
arriving at a {\it unique} final position and velocity for a given particle. 
In the latter one does not always have {\it uniqueness}.
This has been one of the shortcomings of reconstruction,
which was consequently taken to be an ill-posed problem.

In this work, we report on a new method of reconstruction which
guarantees uniqueness (Frisch et al. 2002). 
In Section II, we review the previous works on
reconstruction. We describe our mathematical formulation of
the reconstruction problem
in Section III and show that the cosmological
reconstruction problem, under our hypotheses, is an example of assignment
problems in optimisation theory.
In Section IV, we describe the numerical algorithms to solve the assignment
problem.
In Section V, we test our reconstruction method with numerical N-body
simulation both in real and redshift spaces. 
Section VI contains our conclusions. 

%%%%%%%%%%%%%%%%%%%%%%%%%%%%%%%%%%%%%%%%%%%%%%%%
\section{A brief review of Variational and Eulerian and Lagrangian 
approaches to reconstruction}
%%%%%%%%%%%%%%%%%%%%%%%%%%%%%%%%%%%%%%%%%%%%%%%%

The history of reconstruction goes back to the work of Peebles 
(Peebles 1989) on 
tracing the orbits of the members of the local
group. In his approach, reconstruction was solved as a variational
problem. Instead of solving the equations of motion, one searches
for the stationary points of the corresponding Euler-Lagrange action.
The action in the comoving coordinates is (Peebles 1980)
\be
S=
\int_{0}^{t_0} dt\,\left[
{m_i a^2 \dot{\bf x}_i^2\over 2}
-{G m_i m_j\over a\vert {\bf x}_i-{\bf x}_j\vert}
+{2\over 3}\pi G\rho_{\rm b} a^2 m_i {\bf x}_i^2\right]\;,
\label{action}
\ee
where summation over repeated indices and $j\not=i$ 
is implied, $t_0$ denotes the present time, 
the path of the $i$th particle with mass $m_i$ is ${\bf x}_i(t)$, 
$\rho_{\rm b}$ is the mean mass density, 
and the present value of the expansion
parameter $a(t)$ is $a_0=a(t_0)=1$.
The equation of motion is obtained by requiring
\be
\delta S=
\int_0^{t_0}
dt
\left[
{\partial {\cal L}\over 
\partial{\bf x}_i}\cdot\delta{\bf x}_i
+{\partial{\cal L}\over
\partial\dot{\bf x}_i}\cdot{\delta\dot{\bf x}_i}
\right]
=0\;,
\ee
which leads to 
\be
\int_0^{t_0} dt \delta{\bf x}_i
\left[
\sum_{j\not=i}{G m_i m_j ({\bf x}_i-{\bf x}_j)
\over a \vert {\bf x}_i-{\bf x}_j\vert^3}
+{4\over 3}\pi G\rho_{\rm b} a^2m_i{\bf x}_i
- 
m_i{\partial\over\partial t}(a^2\dot{\bf x}_i^2)
\right]
-m_i
\left[a^2\dot{\bf x}_i\cdot\delta{\bf x}_i\right]^{t_0}_0
=0\;.
\label{umixed}
\ee
The boundary conditions
\ba
\delta{\bf x}_i&=&0 \qquad {\rm at} \qquad t=t_0\nonumber\\
a^2\dot{\bf x}_i&=&0 \qquad {\rm as} \qquad t\rightarrow 0
\label{mixed}
\ea 
would then eliminate the boundary terms in (\ref{umixed}).

%%which would subsequently lead to the equation of motion
%%\be
%%{G m_i m_j ({\bf x}_i-{\bf x}_j)
%%\over a \vert {\bf x}_i-{\bf x}_j\vert^3}
%%+{4\over 3}\pi G\rho a^2m_i{\bf x}_i
%%-m_i{\partial\over\partial t}(a^2\dot{\bf x}^2)=0
%%\ee

The components $\alpha=1,2,3$ of the orbit of the $i$th particle are modeled as
\be
x_i^\alpha=x_i^\alpha(t_0)+\sum_n C_{i,n}^\alpha f_n(t)\;.
\label{coef}
\ee
The functions $f_n$ are
normally convenient functions of the scale factor $a$
and should satisfy the boundary conditions (\ref{mixed}).
Initially, trial functions such as
$
f_n=a^n(1-a)
$
or
$
f_n={\rm sin}\left({n\pi a/ 2}\right)
$
with $f_0={\rm cos}\left({\pi a/ 2}\right)$ were used.
The coefficients $C_{i,n}$ are then found 
by substituting expression (\ref{coef}) in 
the action (\ref{action}) and finding the stationary points. 
That is for physical trajectories
\be
{\partial S\over \partial C_{i,n}}=0\;,
\ee
leading to
\be
m_i\int_0^{t_0} 
dt f_n(t)\left[ -{d\over dt} a^2{dx_i^\alpha\over dt}
+{4\over 3}\pi G\rho_{\rm b} a^2 x_i^\alpha(t)\right]=
-{G\over a}\int^{t_0}_0 dt \sum_j m_j {x_j^\alpha-x_i^\alpha\over
\vert {\bf x}_i-{\bf x}_j\vert}\;.
\label{actionmin}
\ee 

\begin{figure}
\centerline{
        \vspace{-0.8cm}
        \epsfxsize=0.47\textwidth
        \epsfbox{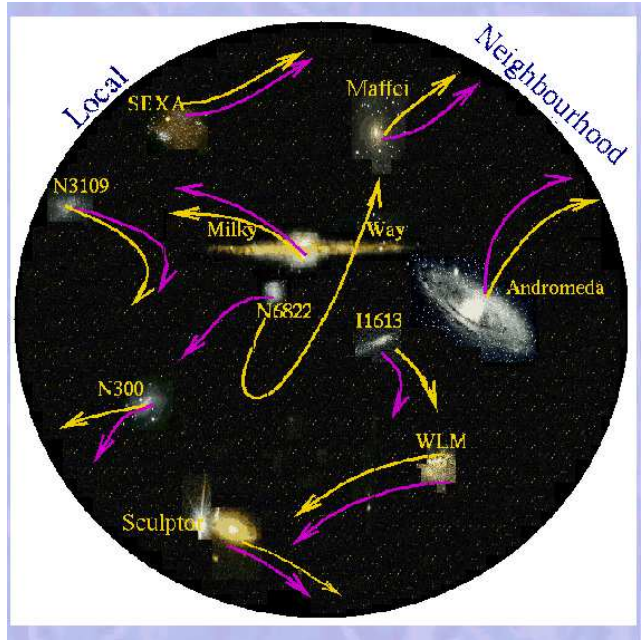}
           }
\vspace{1.2cm}
\caption
{A schematic demonstration of Peebles' reconstruction of the trajectories 
of the members of the local neighbourhood using a variational 
approach based on the
minimisation of Euler--Lagrange action. 
The arrows go back in time, starting from the present 
and pointing towards the initial positions of the sources.
In most cases there is more than one
allowed trajectory due to orbit crossing (closely related to the 
multistreaming of the underlying
dark matter fluid). The pink (darker) orbits correspond to taking the 
minimum of the action whereas the yellow (brighter) orbits were obtained by 
taking the saddle-point solution. 
Of particular interest is the orbit of N6822 which in the former solution is
on its first approach towards us and in the second solution is in its 
passing orbit.
A better agreement between the evaluated and observed velocities was shown to
correspond to the saddle point solution.}
\label{peebles}
\end{figure}

In his first work, Peebles (1989) considered only the 
minimum of the action, while reconstructing 
the trajectories of the galaxies back in time.
In a low-density Universe, assuming a linear bias, 
the predicted velocities agreed with the observed ones
for most galaxies of the local group but failed with a 
large discrepancy for the remaining members.
Later on, it was found that when the trajectories corresponding to the
saddle-point of the action were taken instead of the minimum, much better 
an agreement between predicted
and observed velocities was obtained, for almost all the galaxies 
in the local group (see Fig.\ \ref{peebles}). 
Thus, by adjusting the orbits until the 
predicted and observed velocities agreed, reasonable bounds on cosmological 
parameters were found (Peebles 1989, 1990). 
%%%%%%%%%%%%%%%%%%%% 
%   FIGURE ONE
%%%%%%%%%%%%%%%%%%%%
%%%%%%%%%%%%%%%%%%%%%%

Although rather successful when applied to catalogues such as NGB
(Tully 1988), reconstruction with such an aim, namely establishing
bounds on cosmological parameters using measured peculiar velocities, could not be
applied to larger galaxy redshift surveys, containing
hundreds of thousands of galaxies for the majority 
of which the peculiar velocities are unknown. 
For large datasets, it is not possible to use the velocities, to choose the right
orbit from the many which are all physically possible. In order to resolve 
the problem of multiple solutions (the existence of many mathematically possible 
orbits) one normally had to do significant smoothing and then try the
computationally costly 
reconstruction using Peebles variational approach (Shaya et al. 1995,
Branchini et al. 2001). However, one was still
not guaranteed to have chosen the right orbit (see Hjorteland 1999 for a
review of the action variational method).

The multiple solution can be caused by various factors. For example, the
discretisation in the numerical integrations (\ref{actionmin}) can 
produce spurious valleys in
the landscape of the Euler--Lagrange action. However, even overcoming all these
numerical artifacts one is still not guaranteed uniqueness.
There is a genuine physical reason for the lack of uniqueness which is
often referred to in cosmology as {\it multistreaming}. Cold 
dark matter (CDM) is a
collisionless fluid with no velocity dispersion. An important feature that
arises
in the course of the evolution of a self-gravitating CDM Universe is the formation
of multistream regions 
on which the velocity field is non-unique. These regions are 
bounded by caustics at which the
density is divergent. At a multistream point where velocity is multiple-valued, a
particle can have many different mathematically viable trajectories each of which
would correspond to a different stationary point of 
the Euler--Lagrange action which is
no-longer {\it convex}. 

In addition to the variational approach to reconstruction, 
there is the well-known POTENT reconstruction of the three-dimensional
velocity field from its radial component (Dekel et al. 1990).
The POTENT method is the non-iterative Eulerian version of the original
iterative Lagrangian method (Bertschinger and Dekel 1989) and assumes
potential flow in Eulerian space. POTENT finds the velocity 
potential field by
integrating the radial components of the velocities 
along the line of sight. Since this is an Eulerian method, its 
validity does not extend much beyond the linear Eulerian regime.

In following works, with the use of POTENT-reconstructed velocity field or
the density field obtained from the analysis of the redshift surveys, the
gravitational potential field was also reconstructed 
(Nusser and Dekel 1992). The gravitational potential was
then traced back in time using the so-called Zel'dovich--Bernoulli
equation which combines the Zel'dovich 
approximation with the Euler momentum
conservation equation.
Later on, it was further shown (Gramann 1993) that the initial gravitational 
potential is more accurately recovered using the Zel'dovich-continuity
equation which combines the Zel'dovich approximation with the mass
continuity equation (Nusser et al. 1991). The 
Eulerian formulations of Zel'dovich approximation was 
also extended to second-order and it was found that
this extension does not give significant improvement 
in the reconstruction (Monaco and Efstathiou 1999, see Sahni and 
Coles 1995 for various approximation schemes).

However, these procedures are valid only as long as the density
fluctuations are in the Eulerian linear or quasi-linear regimes 
(defined by $\vert(\rho-\rho_{\rm b})/\rho_{\rm b}\vert\le 1$). 
They do not robustly recover the initial 
density in regions of high density
when the present-day structures are highly nonlinear 
($\vert(\rho-\rho_{\rm b})/\rho_{\rm b}\vert \gg 1$). Therefore, they 
require that the final density field be 
smoothed quite heavily to remove
any severe nonlinearities, before the dynamical evolution equations 
are integrated backward in time.

Here, we describe a new method of 
reconstruction (Frisch et al. 2002) which not only
overcomes the problem of nonuniqueness encountered in 
the least-action-based methods but also remains 
valid well beyond the linear
Eulerian approximations employed in many other popular
reconstruction methods, a few of
which have been mentioned previously. 

%%%%%%%%%%%%%%%%%%%%%%%%%%%%%%%%%%%%%%%%%%%%%%%%%
\section{Monge--Amp\`ere--Kantorovich reconstruction}
%%%%%%%%%%%%%%%%%%%%%%%%%%%%%%%%%%%%%%%%%%%%%%%%%

Thus, reconstruction is a well-posed problem for as long as we avoid
multistream regions. The mathematical formulation of this problem is as
follows (Frisch et al. 2002). Unlike most of the previous works 
on reconstruction where one studies the
Euler-Lagrange action, we start from a constraint equation, namely the mass
conservation,
\be
\rho({\bf x})d{\bf x}=\rho_0({\bf q}) d{\bf q} \;,
\qquad
\ee
where  $\rho_0({\bf q})$ is the 
density at the initial (or Lagrangian) position, ${\bf q}$, and 
$\rho({\bf x})$ is the density at the 
present (or Eulerian) position, ${\bf x}$, of the fluid
element. The above mass conservation equation can be rearranged in the
following form
\be
{\rm det}\left[{\partial q_i\over \partial x_j}\right]=
{\rho({\bf x})\over \rho_0({\bf q})}\;,
\label{det}
\ee
where ${\rm det}$ stands for determinant and $\rho_0({\bf q})$ 
is constant. The right-hand-side of the above
expression is basically given by our boundary conditions: 
the final positions of the particles are known
and the initial distribution is homogeneous, $\rho_0({\bf q}) ={\rm const}$. 
To solve the equation, we make the following two hypotheses:
first that the Lagrangian map (${\bf q}\rightarrow{\bf x}$) is 
the {\it gradient} of a
potential $\Phi$ and second that the potential $\Phi$ is {\it convex} . That is
\be
{\bf x}({\bf q},t)=\nabla_q\Phi({\bf q},t)\;.
\ee
The convexity guarantees that a single Lagrangian position corresponds to a 
single Eulerian position, {\it i.e.}, there has
been no multistreaming
\footnote{ The gradient condition has been used in previous works
 (Bertschinger and Dekel 1989) on the reconstruction of the peculiar
 velocities of the galaxies using linear Lagrangian theory.}.
These assumptions imply that the inverse map ${\bf
 x}\rightarrow {\bf q}$ has also a potential representation
\be
{\bf q}={\bf \nabla}_{\bf x}\Theta ({\bf x},t)\;,
\ee
where the potential $\Theta({\bf x})$ is also a convex function
and is related to $\Phi({\bf x})$ by the Legendre--Fenchel transform
(e.g. Arnold 1978)
\be
\Theta({\bf x})=\max_{\bf q}\left[
{\bf q}\cdot{\bf x}-\Phi({\bf q})\right]\qquad;\qquad
\Phi({\bf q})=
\max_{\bf x}\left[{\bf x}\cdot{\bf q}-\Theta ({\bf x})\right]\;.
\ee
The inverse map is now substituted in (\ref{det}) yielding
\be
{\rm det}\left[{\partial^2 \Theta({\bf x},t)
\over \partial x_i\partial x_j}\right]=
{\rho({\bf x})\over \rho_0({\bf q})}\;,
\label{ma}
\ee
which is the well-known Monge--Amp\`ere equation. 
The solution to this
222 years old problem has recently been discovered
(Brenier 1987, Benamou and Brenier 2000) when
it was realised that the map generated by the solution to the Monge--Amp\`ere
equation is the unique solution to an optimisation problem.
This is the Monge--Kantorovich mass transportation problem, in which one
seeks the map ${\bf x}\rightarrow {\bf q}$ which minimises the quadratic 
{\it cost} function
\be
I=\int_{\bf q} \rho_0({\bf q})\vert{\bf x}-{\bf q}\vert^2 d^3q=
\int_{\bf x} \rho({\bf x})\vert{\bf x}-{\bf q}\vert^2 d^3x \;. 
\label{cost}
\ee
A sketch of the proof is as follows. A small variation in 
the cost function yields
\be
\delta I=\int_{\bf x} \left[2\rho({\bf x})({\bf x}-{\bf q})
\cdot \delta {\bf x}\right] d^3x\;,
\ee
which must be supplemented by the condition
\be
{\bf\nabla}_{\bf x}\cdot\left(\rho({\bf x})\delta{\bf x}\right)=0\;,
\ee
which expresses the constraint that the Eulerian density remains unchanged. 
The vanishing of $\delta I$ should then hold for all ${\bf x}-{\bf q}$
which are orthogonal (in $L^2$) to functions of zero divergence. These are
clearly gradients. Hence ${\bf x}-{\bf q}({\bf x})$ and thus ${\bf q}(\bf x)$
is a gradient of a function of ${\bf x}$.

\begin{figure}
\centerline{
        \vspace{-0.8cm}
        \epsfxsize=0.47\textwidth
        \epsfbox{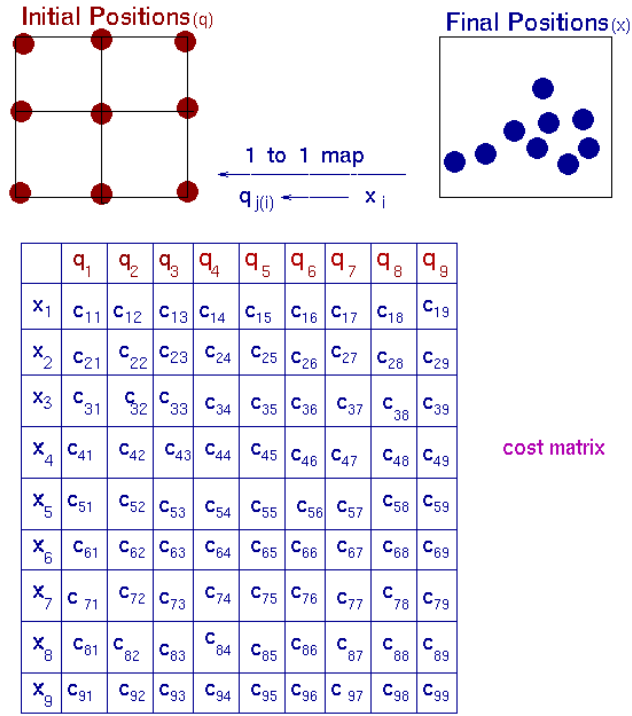}
           }
\vspace{1.1cm}
%{\Large
%$C_{32}$=The cost of getting from $q_2$ to $x_3=(q_2-x_3)^2$
%}
\caption
{
Solving the reconstruction problem as an assignment problem. An example of a
system of $N=9$ particles is sketched. The cost matrix is shown. 
For example the entry ${\rm C}_{32}$ is the cost of getting from
the Lagrangian position ${\bf q}_2$ to the Eulerian position
${\bf x}_3$. The cost of this transport is 
${\rm C}_{32}=({\bf q}_2-{\bf x}_3)^2$. The total cost, 
${\rm C_T}$ of a given permutation is the sum 
of all such costs which are chosen for that permutation. 
That is ${\rm C_T}=\sum_i {\rm C}_{ij(i)}$. 
Generally, there are $N!$ possible permutations.
Only one of these $N!$ permutations 
is the optimal assignment: ${\rm C_{optimal}}=
\min_{j(i)}({\rm C_T})$.
For this system there are 362,880
different costs possible, each obtained 
by a different permutation.
Algorithms with factorial complexity are clearly impractical even for small
systems. However, assignment algorithms 
have complexities of polynomial degrees. 
}
\label{poobles}
\end{figure}

Discretising the cost (\ref{cost}) into equal mass units yields
\be
I=\min_{j(\cdot)}\left(\sum_{i=1}^N\left({\bf q}_{j(i)}-
{\bf x}_i\right)^2\right)\;.
\label{assign}
\ee
The formulation presented in (\ref{assign}) 
is known as the {\it assignment problem}: given $N$ initial and $N$ final
entries one has to find the permutation which minimises the quadratic
cost function (see Fig.\ \ref{poobles}). 

%%%%%%%%%%%%%%%%%%%%%%%%%%%%%%
%    FIGURE TWO
%%%%%%%%%%%%%%%%%%%%

%%%%%%%%%%%%%%%%%%%%%%%%%%%%%%%%%%%%%%%%%
\section{Solving the assignment problem}
%%%%%%%%%%%%%%%%%%%%%%%%%%%%%%%%%%%%%%%%%

If one were to solve the assignment problem for $N$ particles directly, 
one would need to search among $N!$ possible permutations for the one which has the
minimum cost. However, advanced assignment algorithms exist which
reduce the complexity of the problem from factorial to polynomial 
(so far for our problem as verified by 
numerical tests for $N<20,000$, {\it e.g.}, Burkard and Derigs 1980 algorithm 
has a complexity of $N^{3.6}$ for a constant-volume sampling of particles
and a complexity of $N^{2.8}$ for a constant-density sampling. The 
Bertsekas 1998 auction algorithm has a complexity of $N^{2.2}$ in both cases). 
Before discussing these methods, let us briefly
comment on a class of stochastic algorithms, which do not give uniqueness and
hence should be avoided.

\begin{figure}
\centerline{
        \vspace{-0.8cm}
       \epsfxsize=0.47\textwidth
        \epsfbox{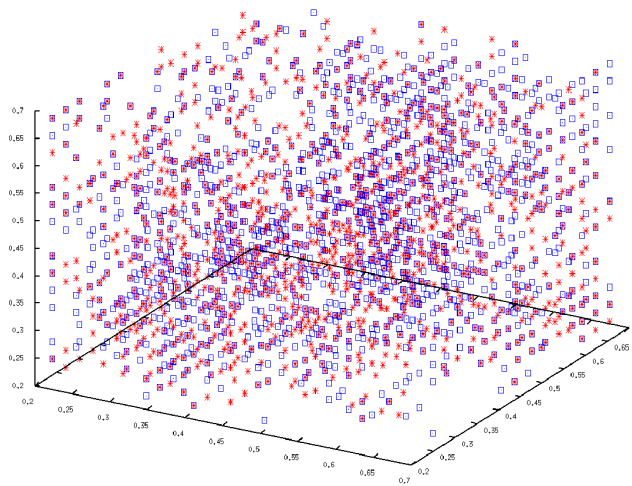}
           }
\vspace{1.0cm}
\caption
{The lack of uniqueness in the results of two runs using a stochastic algorithm
to solve the assignment problems. In the stochastic algorithm, based on
pair-interchange, one searches for, what is frequently referred to in
pure mathematics as, a {\it monotonic map} instead of a true {\it cyclic
monotone map}. The red and blue points (stars and boxes respectively) are the
perfectly-reconstructed Lagrangian positions using the same stochastic code with two
different random seeds. The outputs of the two runs do not coincide, meaning
that the reconstructed Lagrangian positions (and hence peculiar velocities)
depend on the initial random assignment and on the random pair
interchange. The lack of uniqueness in this case is superficial and a 
shortcoming of the chosen numerical method. Such difficulties do not
arise when deterministic algorithms are used to solve 
the assignment problem.}
\label{exactpizaseed1-exactpizaseed2-2}
\end{figure}

In the PIZA method of solving the assignment problem 
(Croft \& Gazta\~naga 1997), initially a random
pairing between $N$ Lagrangian and 
$N$ Eulerian coordinates is made. Starting from this initial random one-to-one
assignment, subsequently a pair (corresponding to two Eulerian and 
two Lagrangian positions) is chosen at random. 
For example, let us consider the randomly-selected pair 
${\bf x_1}$ and ${\bf x_2}$ which have been assigned in the initial random
assignment to ${\bf q}_1$ and ${\bf q}_2$ respectively. Next one swaps
their Lagrangian coordinates and assigns ${\bf x_1}$ to
${\bf q}_2$ and ${\bf x_2}$ to ${\bf q}_1$ in this example. If
\be
\left[({\bf x}_1-{\bf q}_1)^2+({\bf x}_2-{\bf q}_2)^2\right] >
\left[({\bf x}_1-{\bf q}_2)^2+({\bf x}_2-{\bf q}_1)^2 \right]\;,
\ee
then one swaps the Lagrangian positions, otherwise, one keeps 
the original assignment. This process is repeated until one is convinced
that a lower cost cannot be achieved. However, in 
this manner there is no guarantee that the optimal 
assignment has been achieved and the true minimum cost has been found.
Moreover, there is a possibility of deadlock when the cost
can be decreased only by a simultaneous interchange of three or more
particles, while the PIZA algorithm reports a spurious minimum of the
cost with respect to two-particle interchanges.
Results obtained in this way depend strongly on the 
choice of initial random assignment and on
the random selection of the pairs and suffer severely from the lack of
uniqueness (see Fig.\ \ref{exactpizaseed1-exactpizaseed2-2}).

\begin{figure}
\centerline{
        \vspace{-0.8cm}
        \epsfxsize=4. true in
        \epsfbox{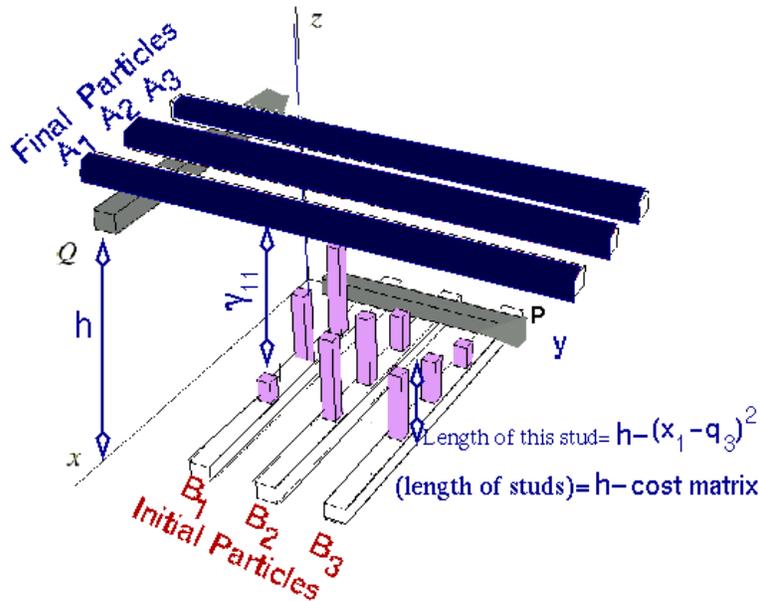}
           }
\vspace{0.8cm}
\caption{  
H\'enon's mechanical device which solves the assignment problem. The device
acts as an analog computer. The rows $A_i$ remain parallel to the $y$ axis,
 are constrained to move
vertically and have positive weights. The columns $B_j$ remain 
parallel to the $x$ axis,
can only move vertically and have negative weights. Vertical 
studs are placed on the columns, in such a
way that each stud enforces a minimal distance between row $A_i$ 
and column $B_j$. Initially all rods are kept at fixed positions by stops $P$
and $Q$. Then the rods are released by removing the stops $Q$ and $P$ and the
system starts evolving. Rows go down and columns go up and aggregates of rows
and columns are made. Thus a complex evolution takes place. The rules for the
formation of aggregates are demonstrated by a simple example in 
the figure that follows.}.
\label{henon1}
\end{figure}

In spite of these problems associated with PIZA-type 
algorithms in solving the
assignment problem, it has been
shown (Narayanan and Croft 1999) in a point-by-point comparison
of the reconstructed with the true initial density field from 
a N-body simulation, that PIZA performs better 
than other methods such as those based on a Eulerian
formulation of the Zel'dovich 
approximation (discussed at the end of Section II).
%%%%%%%%%%%%%%%%%%%%%%%%%%%%%%%%%%%%%%%%%%%%%%%%%%%%%%%%%%%%%%%%%%%%%%%%%%%%%%%%%%%
%  FIGURE THREE
%%%%%%%%%%%%%%%%%%%%%%%%%%%%%%%%%%%%%%%%%%%%%%%%%%%%%%%%%%%%%%%%%%%%%%%%%%%%%%%%%%%%
%%%%%%%%%%%%%%%%%%%%%%%%%%%%%%%%%%%%%%%%%%%%%%%%%%%%%%%%%%%%%%%%%%%%%%%%%%%%%
There are various deterministic algorithms which guarantee that the
optimal assignment is found. An example is an algorithm written by
H\'enon (H\'enon 1995), demonstrated in Fig.\ \ref{henon1}.
In this approach, a simple mechanical device is defined and then simulated 
which solves the 
assignment problem. The device acts as an {\it analog computer}: the numbers 
entering the problem are represented by physical quantities and the equations 
are replaced by physical laws.
 
One starts with $N$ columns $B$s (which represent the Lagrangian positions 
of the particles) 
and $N$ rows, $A$s (which represent the Eulerian 
positions of the particles). On each row
there are $N$ studs whose lengths are directly related to the
distances between that row and each column. The columns are given negative weights
of $-1$ and act as floats and the rows are given weights of $1$.

The potential energy of the system shown in Fig.\ \ref{henon1}, 
within an additive constant, is
\be
U=\sum_i \alpha_i-\sum_j \beta_j\;,
\label{pe}
\ee
where $\alpha_i$ is the height of the row $A_i$ and 
$\beta_j$ is the height of the column $B_j$, since all rows and columns
have the same weight ($1$ and $-1$ respectively).

\begin{figure}[htb]
\centerline{
        \vspace{-0.8cm}
        \epsfxsize=5 true in
  \epsfbox{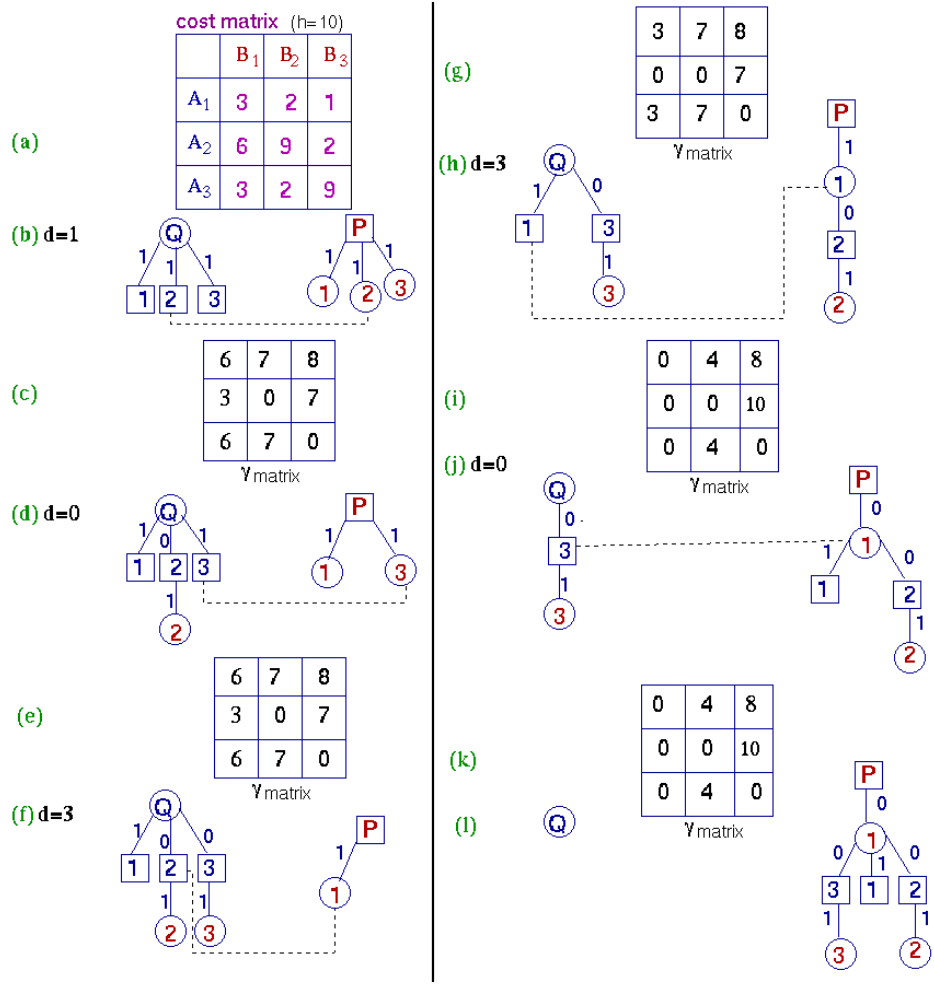}
           }
\vspace{1.cm}
\caption{  
A step-by-step (a to l) progress of H\'enon's algorithm on a simple
example with three initial and final positions (columns and rows) is shown.
The table on the top left shows the values of the costs (distances between 
rows and columns). When executing the algorithm by hand, it is convenient 
to keep track of the distances between the rows and the studs, {\it i.e.} the
quantities
$\gamma_{ij}=\alpha_i-\beta_j-c_{ij}$ where $\alpha_i$ is the variable height
of the row $A_i$ and $\beta_j$ is the variable height of the column $B_j$.
A graph of the initial state is made with the first contact being made
between the second row and second column which have the largest cost (note
that the code originally written by H\'enon, in fact, finds the maximum
cost. For our purpose of finding the minimum cost, one can just simply 
subtract the matrix elements $c_{ij}$ from a large number). Thus, at this point
the entries in the $\gamma$ matrix change since now the second row and column
are in contact and hence $c_{22}=0$. Obviously the distances between all other
rows and columns should also be modified. The second matrix shows the new
distances and automatically a contact is made between third row and 
third column whose separation is now also zero. Since the second and third 
rows cannot move now, the next contact is made between the first column and 
the second row and the break occurs where the total force on the branch is 
weakest.
Since this is where the second row meets the support $Q$, part of the $Q$ tree
is captured by the $P$ tree, as demonstrated.
The next contact can now only be made between 
the first row and the first column and the
break occurs at the weakest branch. The equilibrium position is now reached
where each row is supported by one column. In the final optimal assignment
column 1 is attached to row 1, column 2 goes to row 2 and 
column 3 is associated to row 3.
For this simple exercise one can
easily see that this procedure achieves the maximum cost (which in 
this example is $21$).}
\label{henon2}
\end{figure}

Initially, all rods are maintained at a fixed position by two additional rods
$P$ and $Q$ with the row $A_i$ 
above column $B_j$, so that there is no contact
between the rows and the studs. Next, the rods are released by removing the
holds $P$ and $Q$ and the system starts evolving: rows go down 
and columns go
up and the contacts are made with the studs. Aggregates of rows and columns
are progressively formed. As new contacts are made, these aggregates are
modified and thus a complicated evolution 
follows which is graphically demonstrated with a
simple example in Fig.\ \ref{henon2}. One can then show that an
equilibrium where the potential energy (\ref{pe}) of the system is minimum 
will be reached after a {\it finite time}. It 
can then be shown (H\'enon 1995) that if the
system is in equilibrium and the column $B_j$ is in contact with row $A_i$
then the force $f_{ij}$ is the optimal solution of the assignment problem. 
When $f_{ij}$ is equal to one there is an assignment between rows and columns
and when $f_{ij}$ is equal to zero there is no assignment.
The potential energy of the corresponding equilibrium is 
equivalent to the total {\it cost} of the optimal solution.

%%%%%%%%%%%%%%%%%%%%
%  FIGURE FOUR
%%%%%%%%%%%%%%%%%%%%
%%%%%%%%%%%%%%%%%%%%%%%%%%%

%%%%%%%%%%%%%%%%%%%%%%%%%%%%%%%%%%%%%%%%%%%%%%%%%%%%%%%%
\section{Test of Monge--Amp\`ere--Kantorovich (MAK) reconstruction with 
N-body simulations}
%%%%%%%%%%%%%%%%%%%%%%%%%%%%%%%%%%%%%%%%%%%%%%%%%%%%%%%%

Thus, in our reconstruction method, the initial
positions of the particles are uniquely found by solving the assignment
problem. 
This result was based on our reconstruction hypothesis.  We could test the
validity of our hypothesis by direct comparison with numerical N-body 
simulations 
which is what we shall demonstrate later in this section. However, it is
worth commenting briefly on the theoretical, observational and numerical 
justifications for our hypothesis. 
It is well-known that the Zel'dovich approximation 
(Zel'dovich 1970) works well in describing
the large-scale structure of the Universe. In the Zel'dovich approximation
particles move with their initial velocities on inertial trajectories
in appropriately redefined coordinates. It is also known that the
irrotationality of
the particle displacements (expressed in Lagrangian coordinates) remains valid
even at the second-order in the Lagrangian perturbation theory (Moutarde et
al. 1991; Buchert 1992; Munshi et al. 1994; Catelan 1995). This provides the
theoretical motivation for our first hypothesis. The lack of severe multistream
regions is confirmed by the geometrical properties
of the cosmological structures. In the presence of significant multistreaming
such as the one that occurs in Zel'dovich approximation, one could not expect
the formation of long-lived structures such as filaments and walls which
is observed in
numerical
N-body simulations. In
the presence of significant multistreaming these structures would form and
would smear out and disappear rapidly. This is not backed by numerical
simulations. The latter show the formation of shock-like structures well-described
by Burgers model (a model of shock formation and evolution in a
compressible fluid), which has been extensively used to describe the large-scale
structure of the Universe (Gurbatov \& Saichev 1984; 
Gurbatov, Saichev \& Shandarin 1989; Shandarin \& Zel'dovich
1989; Vergassola et al. 1994). The success of Burgers model indicates that a
viscosity-generating mechanism operates at 
small scales in collisionless systems resulting in
the formation of shock-like structures rather than caustic-like structures.

\begin{figure}
\centerline{
        \vspace{-0.8cm}
        \epsfxsize=0.49\textwidth
        \epsfbox{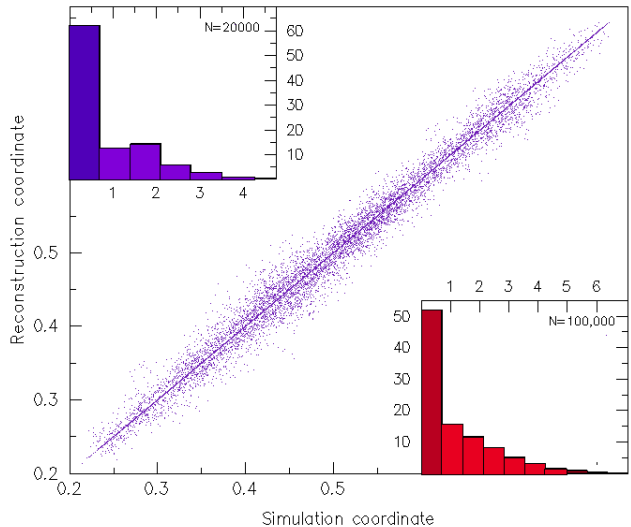}
           }
\vspace{1.0cm}
\caption
{\small Test of MAK reconstruction of the Lagrangian positions, 
using a $\Lambda$CDM simulation of $128^3$ particles in a box of
size $200 ~{\rm Mpc}^3/h^3$. In the scatter plot, the dots near the diagonal
are a scatter plot of reconstructed initial points versus simulation initial
points for a grid of size $6$ Mpc/h with about 20,000 points. The scatter
diagram uses a {\it quasi-periodic projection} coordinate
${\tilde {\bf q}}\equiv (q_x+\sqrt 2 q_y+{\sqrt 3} q_z)/(1+
\sqrt 2+\sqrt 3)$ which guarantees a one-to-one correspondence between $\tilde
{\bf q}$ values and points on the regular Lagrangian grid. The upper left
inset is a histogram (by percentage) of distances in reconstruction mesh 
units between such points; the first bin corresponds to perfect
reconstruction;
the lower-inset is a similar histogram for reconstruction on a finer grid 
of about $3$ Mpc/h using $100,000$ points. With the $6$ Mpc/h grid $62\%$
of the points, 
and with $3$ Mpc/h grid more than $50\%$ of the points, are assigned perfectly.
}
\label{pobles}
\end{figure}

In spite of this evidence in support of our reconstruction
hypothesis, one needs to test it before applying it
to real galaxy catalogues.
We have tested our reconstruction against numerical N-body simulation. We ran
a $\Lambda$CDM simulation of $128^3$ dark matter particles, using the
adaptive P$^3$M code HYDRA (Couchman et al. 1995). Our cosmological 
parameters are $\Omega_m=0.3,
\Omega_\lambda=0.7, h=0.65, \sigma_8=0.9$ and a box size of
$200$Mpc/h. We took two sparse samples 
of $17,178$ and $100,000$ particles corresponding
to grids of $6$Mpc and $3$Mpc respectively, at $z=0$ and
placed them initially on a uniform grid. 
For the $17,178$ particle reconstruction, 
the sparse sampling was done by first selecting a
subset of $32^3$ particles on a regular Eulerian grid 
and then selecting all the particles which remained inside a 
spherical cut of radius half the size 
of the simulation box.
The sparse sample of $100,000$ points was made randomly. 
An assignment algorithm, similar
to that of H\'enon described in the previous section, is used
to find the correspondence between the Eulerian and the Lagrangian 
positions. The results of our test are shown in Fig.\ \ref{pobles}. 

%%%%%%%%%%%%%%%%%%%%%%%%%%%%%%
%  FIGURE 5
%%%%%%%%%%%%%%%%%%%%

\begin{figure}
\centerline{
        \vspace{-0.8cm}
       \epsfxsize=0.6\textwidth
        \epsfbox{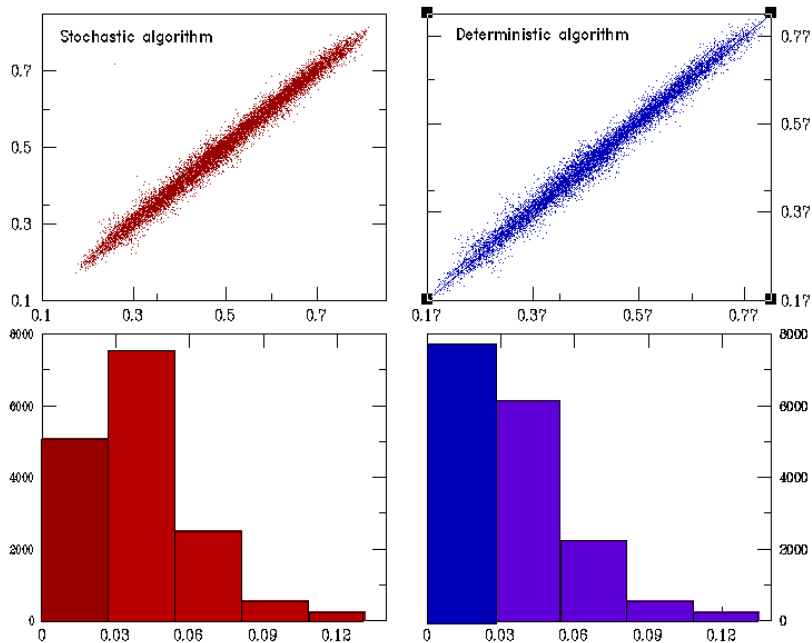}
           }
\vspace{1.3cm}
\caption
{
A comparison of a PIZA-type, stochastic method, versus our MAK,
deterministic method of solving the assignment problem is shown. 
The outputs of the simulation 
used for Fig.\ \ref{pobles} have also been used
here. In the stochastic method, almost one billion pair interchange are
made. At which point the cost-reduction cannot be reduced anymore.
However, even with such a large number of interchanges, the number of exactly
reconstructed points is well below that achieved by MAK. 
The upper sets are the scatter plots of reconstructed versus simulation
Lagrangian positions for PIZA (left top set) and for MAK (right top set).
The lower sets are histograms of the distances between the reconstructed
Lagrangian positions and those given by the simulation. The bin sizes are
taken to be
slightly less than one mesh. Hence, all the points in the first (darker) bin
correspond to those which have been exactly reconstructed.
Using
PIZA, we achieve a maximum of about $5000$ out of the $17,178$ points 
exactly-reconstructed positions
whereas with MAK this number reaches almost $8000$ out of $17,178$ points. Note 
that, for the sole purpose of comparing MAK with PIZA it is not
necessary to account for periodicity corrections, which would improve both
performances equally.
Accounting for the periodicity improves exactly-reconstructed MAK positions
to almost $11,000$ points out of $17,178$ points used for reconstruction, as
shown in the upper inset of Fig.\ \ref{pobles}.
Starting with an initial random cost of about 200 million (Mpc/h)$^2$ 
($5000$ in our box
unit which runs from  zero to one), after one billion pair
interchange, a minimum cost of about 1,960,000 (Mpc/h)$^2$ ($49$ in our box unit) 
is achieved. Continuing to 
run the code on a
2\, GHz DEC alpha workstation, consuming almost a week of CPU time, does not
reduce the cost any further. With the MAK algorithm, the minimum
cost is achieved, on the same machine, in a few minutes. The cost for 
MAK is 1,836,000 (Mpc/h)$^2$ ($45.9$ in box units) which is
significantly lower than the minimum cost of PIZA. Considering that the average
particle displacement in one Hubble time is about $10$ Mpc/h (about $1/20$ of
the box size) this discrepancy between MAK and PIZA costs 
is rather significant.
}
\label{mak1-piza_nonmonotonic_seed1}
\end{figure}

It is instructive to compare our results presented in Fig.\ \ref{pobles}
with those obtained by a stochastic PIZA-type algorithms.
Fig.\ \ref{mak1-piza_nonmonotonic_seed1} demonstrates such a comparison.
We see that not only the stochastic algorithms do not guarantee uniqueness,
as demonstrated in Fig.\ \ref{exactpizaseed1-exactpizaseed2-2},
but also the number of exactly reconstructed points by these algorithms are
much below those obtained by our deterministic algorithm.

%%%%%%%%%%%%%%%%%%%%%%%%%%%%%%%%%%%%%%%%%%%%%%%%%%%%%%%%%%%%%%
%%%%%%%%%%%%%%%%%%%%%%%%%%%%%%%%%%%%%%%%%%%%%%%%%%%%%%%%%%%%%%%
When reconstructing from observational data, in redshift space, the galaxies
positions are displaced radially by an amount proportional to the radial
component of the peculiar velocity in the line of sight.
Thus for real catalogues, the cost minimisation need be performed using redshift
space coordinates (Valentine et al. 2000). 
The redshift position of a galaxy is given by
\be
{\bf s}={\bf x}+\hat{\bf s}\left({\dot {\bf x}\over H}\cdot 
\hat{\bf s}\right)\;,
\label{red}
\ee
where under our assumption of irrotationality and 
convexity of the Lagrangian map and to first order in 
Lagrangian perturbation theory,
\be
\dot{\bf x}=\beta H ({\bf x}-{\bf q})
\ee
and
$
\beta={f(\Omega)/ b}
$
with $f(\Omega)=d{\rm ln}\delta/d{\rm ln}a=\Omega_{\rm m}^{0.6}$ and 
the bias factor is
$\delta_{\rm galaxies}=b\delta_{\rm mass}$. The quadratic cost function can
then be easily found in terms of redshift space coordinates as follows
\be
({\bf x}-{\bf q})^2=
({\bf s}-{\bf q})^2-
{\beta(2+\beta)\over (1+\beta)^2}
\left(({\bf s}-{\bf q})\cdot
\hat{\bf s}\right)^2\;.
\label{costred}
\ee
Using MAK, based now on the modified cost (\ref{assign}) we 
can then find the corresponding
${\bf q}$ for each ${\bf s}$ and hence find the peculiar velocities. Indeed,
using (\ref{red})
\be
{\bf s}-{\bf q}=
{\dot{\bf x}\over H}\left(1+{b\over \Omega_{\rm m}^{0.6}}\right)\;.
\ee
Note that in this approach the motion of the local group itself is
ignored, which however should also be accounted 
for (Taylor \& Valentine 1999).
Furthermore, the effect of catalogue selection function can be 
handled by giving each galaxy an {\it effective-mass} inversely proportional to the 
catalogue selection function (Nusser and Branchini 2000). 
We have also assumed that galaxies are unbiased tracers of the mass
distribution. However, the relative bias between galaxies of different
luminosities and
of different morphological types indicate that galaxies are likely to be
biased tracers of mass distribution (e.g. Loveday et al. 1995).
Biasing can be taken into account in a similar manner to the catalogue
selection function (Nusser and Branchini 2000). The net effect 
of biasing for our scheme is that it changes
the effective-mass associated to the galaxies which can be easily incorporated 
in our cost function (\ref{assign}), by multiplying the quadratic function
inside the sum by the effective-mass $m_i$ associated with each Eulerian position
${\bf x}_i$.

\begin{figure}
\centerline{
        \vspace{-0.3cm}
        \epsfxsize=0.58\textwidth
        \epsfbox{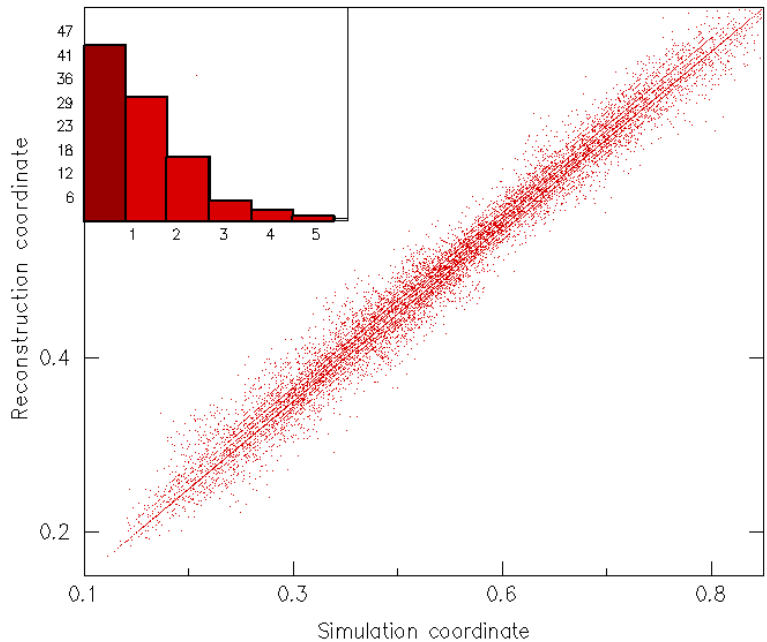}
           }
\vspace{1.0cm}
\caption
{
Reconstruction test in redshift space with the same data as that used for
real-space reconstruction tested in the 
upper left histogram of Fig.\ \ref{pobles}. The velocities are smoothed
over a sphere of radius $2$ Mpc/h. The
observer is taken to be at the centre of the simulation box. 
The Lagrangian reconstructed points are plotted against the simulation
Lagrangian positions using the quasi-periodic projection coordinates used
for the scatter plot of Fig.\ \ref{pobles}. 
The histogram corresponds to the distances between the
reconstructed and the simulation Lagrangian positions 
for each Eulerian position.
The bins of the histogram
are smaller than one mesh and the first one corresponds to 
exactly-reconstructed Lagrangian positions. We obtain 
$43\%$ perfect reconstruction.}
\label{fig3-marseille}
\end{figure}

We have also
performed MAK reconstruction with the redshift-modified cost function 
which led to somewhat degraded results but at the same time provided an 
approximate determination of peculiar velocities 
(see Fig.\ \ref{fig3-marseille}). A more accurate
determination of peculiar velocities can be made using second-order 
Lagrangian perturbation theory (e.g. Catelan 1995 and refs. therein). 

%%%%%%%%%%%%%%%%%%%%
%%%%\begin{figure}
%%%%\centerline{
%%%%        \vspace{-0.3cm}
%%%%        \epsfxsize=0.4\textwidth
 %%%%       \epsfbox{redshift-mak-sim.ps}
%%%%           }
%%%%\vspace{1.0cm}
%%%%\caption
%%%%{
%%%%}
%\label{pebles}
%%%%\end{figure}
%%%%%%%%%%%%%%%%%%%%
%  FIGURE 6
%%%%%%%%%%%%%%%%%%%%
\begin{figure}
\centerline{
        \vspace{-0.3cm}
        \epsfxsize=0.45\textwidth
        \epsfbox{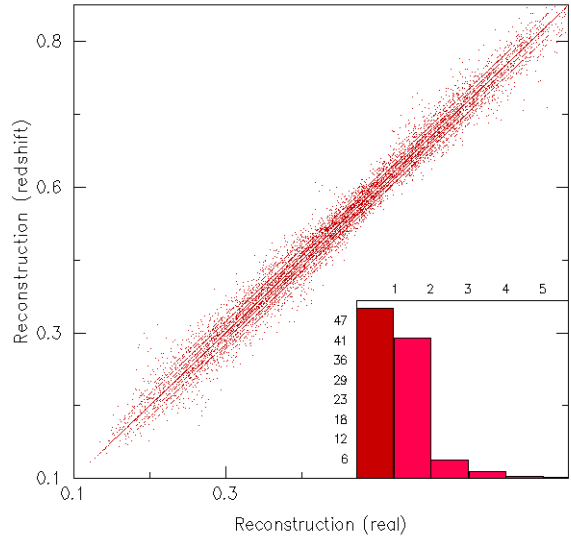}
           }
\vspace{1.0cm}
\caption
{
Comparison of redshift versus real space reconstruction allows a test of
the robustness of MAK reconstruction against systematic errors. 
The same data as those used for 
Fig.\ \ref{fig3-marseille} has been used to obtain the
above result. In both real and
redshift space most of the exactly reconstructed points 
almost $50\%$ of the points, which fall in the 
first bin of the histogram,
correspond to the same particles.
}
\label{redshiftreal}
\end{figure}

We have also compared the redshift-space MAK reconstruction to real space 
MAK reconstruction, which shows that almost $50\%$ of exactly-reconstructed
positions correspond to the same points (see Fig.\ \ref{redshiftreal}).
This test shows that MAK is robust against systematic errors.
%%%%%%%%%%%%%%%%%%%%%%
%COMPARISON OF REDSHIFT VERSUS REAL SPACE RECONSTRUCTION: SYSTEMATIC ERRORS
%%%%%%%%%%%%%%%%%%%%
%%%%%%%%%%%%%%%%%%%%%%
Therefore the question that one asks at this point is: where does
reconstruction perform poorly?
%%%%%%%%%%%%%%%%%%%%%%%%%%%%%%%%%%%%%%%%%%%%%%%%%%%%%%%%%%%
\begin{figure}
\centerline{
        \vspace{-0.3cm}
        \epsfxsize=0.47\textwidth
        \epsfbox{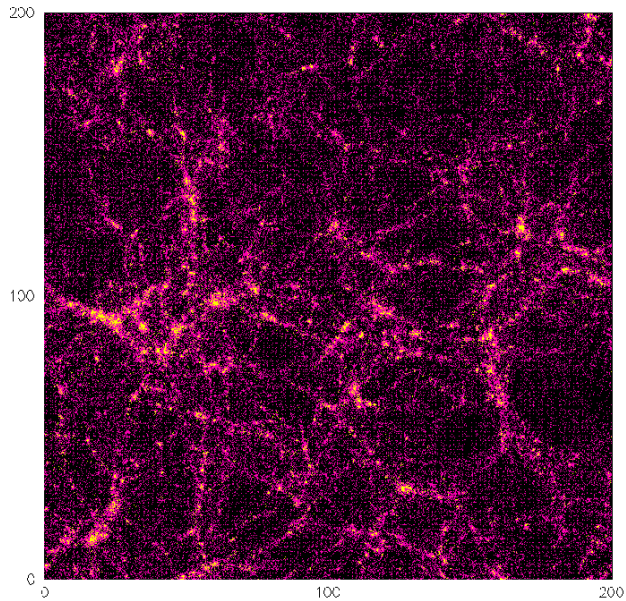}
           }
\vspace{1.0cm}
\caption
{
A thin slice (26 Mpc/h) of the box (of sides 200 Mpc/h) of 
the $ 128^3, \Lambda$CDM simulation
the same as that used for Fig.\ \ref{pobles} is shown.
Out of the $100,000$ points which have been used for reconstruction 
more than half are exactly reconstructed. The
points which fail exact reconstruction are highlighted (yellow, brighter
points).
The plot verifies that these points lie 
mainly in the overdense regions where
we do not expect our reconstruction hypothesis to hold well.
}
\label{failed}
\end{figure}
%%%%%%%%%%%%%%%%%%%%%%%%%%%%%%%%%%%%%%%%%%%%%%%%%%%%%%%%
Fig.\ \ref{failed} highlights where exact reconstruction fails.
We see that exact reconstruction is not achieved in 
particular in the dense regions. Achieving 
reconstruction at small
scales remains a subject of on-going research. As long as multistreaming
effects are unimportant, that is above $\simeq 1$ Mpc, uniqueness of
the reconstruction is guaranteed. Approximate algorithms capturing
such highly nonlinear effects are now being developed.

%%%%%%%%%%%%%%%%%%%%%%%%%%%%%%%%%%%%%%%%%%%%%%%%%%%%%%%%
\section{Conclusion}
%%%%%%%%%%%%%%%%%%%%%%%%%%%%%%%%%%%%%%%%%%%%%%%%%%%%%%%%

An accurate reconstruction allows us to test cosmological models by simulating
the evolution starting from the reconstructed primordial state and comparing
it to observations.
Efficient and unique reconstruction would allows us to 
determine the peculiar velocities of a large number of galaxies, using their
positions in redshift catalogues.
One could use reconstruction to search for 
the presence 
of primordial non-gaussianity and 
examine the self-consistency of cosmological hypotheses such as the choice of
the global cosmological parameters and the assumed biasing scheme. 
By obtaining a point-by-point
reconstruction
of the specific realisation that describes the 
observed patch of our Universe, one could separate the 
universal properties and the influence of
the large-scale environment on the galaxy formation process.

Many previous reconstruction techniques have either suffered
from the lack of uniqueness or limited validity which could 
not be extended beyond the linear Eulerian regime.
Our reconstruction scheme overcomes such shortcomings.

We have shown that under suitable hypotheses, there is a unique
transformation from the present positions 
${\bf x}$ of galaxies (taken as mass tracers) to their respective initial positions
${\bf q}$. Our reconstruction hypothesis first assumes
that the Lagrangian map ${\bf x}\rightarrow {\bf q}$ can be written in terms of
a potential ${\bf x}={\bf\nabla}_{\bf q}\Phi({\bf q})$. This hypothesis 
is supported by
numerical N-body simulations and is valid up to second order Lagrangian
perturbation theory. The second assumption is the
convexity of the potential
$\Phi({\bf q})$, a consequence of which is the absence of multi-streaming:
for almost any Eulerian position, there is a single Lagrangian antecedent.  
As is well-known, the Zel'dovich approximation leads to caustics and
multistreaming. This can be modified
in various ways, for example by the introduction of a mock viscosity 
as is done in the adhesion model (Gurbatov and 
Saichev 1984, Gurbatov et al. 1989, 
Shandarin and Zel'dovich 1994). The latter which leads to shocks
rather than caustics, is known to have a convex 
potential (Vergassola et al. 1994) and to be in a 
better agreement with numerical N-body simulations (Weinberg and Gunn 1990).

By testing our MAK reconstruction against numerical N-body simulations both in
real and redshift spaces, 
we have demonstrated that our
reconstruction hypothesis works well down to the scales 
comparable to the size of collapsed structures, below which the hydrodynamical
description of gravitational structure formation ceases to be meaningful
(at scales larger than $6$ Mpc/h we achieve more than $62\%$ exact
reconstruction). Thus, our reconstruction is rather suitable for 
large-scale galaxy redshift surveys.

We thank E. Branchini, M. H\'enon, J.A. de Freitas Pacheco 
and P. Thomas for discussions and comments. 
U.F thanks the Indo-French Center for the 
Promotion of Advanced Research (IFCPAR 2404-2).
A.S. is supported by A CNRS Henri 
Poincar\'e fellowship and RFBR 02-01-01062.
R.M. is supported by the European Union 
Marie Curie Fellowship HPMF-CT-2002-01532.
%%\vspace{-1.5cm}
%%%%%%%%%%%%%%%%%%%%%%%%%%%%%%
%%%%%%%%%%%%%%%%%%%%%%%%%%%%%%

\end{document}